\documentclass[12pt,a4paper]{article}

  \usepackage{geometry}
  \usepackage{latexsym}
  \usepackage{epsf}
  \usepackage{amssymb}
  \usepackage{graphicx,color}
  \usepackage{amsmath, cite}
  \usepackage{amsmath,amssymb,amsthm}
  \usepackage{verbatim}
  \usepackage{hyperref}
\renewcommand{\d}{\textrm{d}}

\newcommand{\w}{\wedge}

\newcommand{\SU}{\mathop{\rm SU}}

\newcommand{\kper}{k_{\perp}}
\newcommand{\kpar}{k_{\parallel}}

\renewcommand{\Im}{\operatorname{Im}}

\newcommand{\be}{\begin{equation}}
\newcommand{\ee}{\end{equation}}
\newcommand{\beq}{\begin{equation}}
\newcommand{\eeq}{\end{equation}}
\newcommand{\ba}{\begin{eqnarray}}
\newcommand{\ea}{\end{eqnarray}}
\newcommand{\bea}{\begin{eqnarray}}
\newcommand{\eea}{\end{eqnarray}}

\renewcommand{\d}{\textrm{d}}

\begin{document}
\pagestyle{empty}
\numberwithin{equation}{section}
\begin{center}

\begin{Huge}
\textbf{A method to find $\mathcal{N}=1$ AdS$_4$ vacua in type IIB}
\end{Huge}

\vspace{1cm}

\begin{Large}
Gautier Solard
\end{Large}

\vspace{1cm}

\emph{Physics Department, Universit\`a  di Milano-Bicocca, Piazza della Scienza 3,20100 Milano, ITALY}

\emph{INFN, sezione di Milano-Bicocca, Milano, ITALY}

\vspace{1cm}

email : gautier.solard@mib.infn.it
\end{center}

\vspace{1cm}

\begin{abstract}
In this paper, we are looking for $\mathcal{N}=1$, AdS$_4$ sourceless vacua in type IIB. While several examples exist in type IIA , there exists only one example of such vacua in type IIB . Thanks to the framework of generalized geometry we were able to devise a semi-algorithmical method to look for sourceless vacua. We present this method, which can easily be generalized to more complex cases, and give two new vacua in type IIB.
\end{abstract}

\clearpage
\setcounter{page}{1}
\pagestyle{plain}
\tableofcontents

\section*{Introduction}

Compactification to 4-dimensional anti-De Sitter (AdS$_4$) are of relevance to several aspects of string theory. In particular, they are central in the CFT$_3$/AdS$_4$ correspondence. They can also be a first step toward obtaining a De Sitter vacuum if one devise a way to break supersymetry in a controlled way.

In type IIA, several AdS$_4$ vacua have been found without  \cite{Tomasiello:2007eq,Petrini:2009ur,Lust:2009mb,Koerber:2010rn,Guarino:2015jca} or with \cite{Dasgupta:1999ss,Giddings:2001yu,Behrndt:2004km,Derendinger:2004jn,Lust:2004ig,DeWolfe:2005uu,Koerber:2008rx,Caviezel:2008ik,Varela:2015uca} sources (this is a non exhaustive list of examples). On the contrary, in type IIB, there have been far less studies. Some results have been found with sources \cite{Lust:2009zb,Petrini:2013ika,D'Hoker:2007xy,Assel:2011xz} but only one example without sources \cite{Lust:2009mb} (even if the solution is singular in the compactified description). It is to remedy to this state of affairs that we looked for more sourceless vacua in type IIB. This type of vacua also presents two advantages. The first one is, as we already mentioned, their use in the CFT$_3$/AdS$_4$ correspondence. The second one is the validity of such vacua. Indeed, in most known examples with sources, the sources are smeared and one can ask if   this assumption is well-founded. Getting rid of the sources also gets rid of this problem.

In order to find sourceless vacua, we use the pure spinors formalism developed in \cite{Grana:2006kf,Grana:2004bg,Grana:2005sn}. This permits to obtain linear algebraic equations for the SUSY equations. We are left, thanks to the integrability theorem \cite{Lust:2004ig,Gauntlett:2005ww,Koerber:2007hd}, with the Bianchi identities which are quadratic and differential. Since these are not solvable in all generality, one has to devise a way to solve them. Taking inspiration from \cite{Solard:2013fva}, where parts of the quadratic equations were in fact linear and permitted to solve the whole system of quadratic equations, we put in place a semi-algorithmical method to solve the equations. We also had to take care of the differential part which was absent from \cite{Solard:2013fva}. This method can be easily generalized to all type of problems with the same characteristics. Thanks to it, we were able to recover an example of the known sourceless vacuum \cite{Lust:2009mb} and discover two new vacua which are a priori sourceless. A more careful study shows that these solutions are singular and we give for one of these examples a possible interpretation in terms of sources.

\vspace{0.5cm}

This paper is organized as follows. In section \ref{susysect}, we present the supersymmetry conditions in the framework of generalized geometry applied to our specific case. In section \ref{method}, we expose the method to solve the quadratic equations. Finally in section \ref{examples}, we give three examples of vacua, one of them already known that we recover thanks to our method and two new ones.

\section{The supersymmetry conditions}
\label{susysect}

We are interested in $\mathcal{N}=1$ SUSY AdS$_4$ vacua in type IIB theories. That is to say that the manifold the theory lives on is of the type :
 \beq
 \label{10dmetric}
 \d s^2 = e^{2 A} \d s^2_{(4)} + \d s^2_{(6)} \, ,
 \eeq
with A the warp factor.
 As discussed in \cite{Behrndt:2005bv, Lust:2009zb} such solutions are only possible
when the compactification manifold have SU(2) structure group. Let us recall that a manifold  is said to be of SU(2) structure if it admits 
a complex one form $z$,  a real and a holomorphic two-form, $j$ and $\omega$, that are globally defined and satisfy
\begin{subequations}
\bea
&& z \llcorner \bar{z} = 2\,, \qquad   z \llcorner z =  \bar{z} \llcorner \bar{z} = 0\,, \\
&& j \w \omega =0 \,, \\
&&  z \llcorner j  = z \llcorner \omega = 0 \,, \\
&& j \w j = \frac{1}{2} \omega \w \bar{\omega} \,.
\eea
\end{subequations}

\vspace{0.3cm}

In order to study $\mathcal{N}=1$ vacua with non trivial fluxes, it is convenient to use the language of  Generalized Complex Geometry \cite{Hitchin:2004ut,Gualtieri:2003dx}. We will give here a lightning review restricted to our specific case, for some more details, see for example \cite{Petrini:2013ika,Solard:2013fva} and references therein.

The idea is to express the ten-dimensional supersymmetry variations as differential equations on a pair of polyforms defined on the internal manifold. In our case they are 
\bea
\label{Phim}
&& \Phi_-= -\frac{e^A}{8}   z \wedge (\kper e^{ -ij }  + i \kpar \omega ) \, , \\
\label{Phip}
&& \Phi_+=  \frac{ e^A e^{i\theta}}{8}   \, e^{ z \bar{z} /2} (\kpar e^{-ij}  -i\kper \omega ) \, ,
\eea
where $z$, $j$ and $\omega$ are the forms defining the $\SU(2)$ structure, A the warp factor and $\theta$ a free parameter. The parameters $\kpar$ and $\kper$ ($\kpar^2 + \kper^2 = 1$) 
are related to the choice  of structure on the internal manifold.  
When  $\kpar = 0$ and $\kper=1$ the structure is strict $\SU(2)$, while the general case where both $\kpar$ and $\kper$ are non-zero is often referred to as dynamical $\SU(2)$  structure\footnote{When $\kpar = 1$ and $\kper=0$ the internal manifold is said to be of $\SU(3)$ structure. We will not consider
this case here.}.  When $\kpar$ and $\kper$ are non zero and constant, we speak of intermediate SU(2) structure rather than dynamical SU(2) structure \cite{Andriot:2008va}.

\vspace{0.3cm}

As shown in \cite{Grana:2004bg},  for type IIB compactifications to AdS$_4$  the ten-dimensional supersymmetry variations  are equivalent to the following set of equations on the pure spinors $\Phi_\pm$ 
\begin{subequations}
\bea
\label{eqm}
&&(\d - H  \w )(e^{2A - \phi} \Phi_-)  = - 2 \mu e^{A-\phi} {\rm Re } \Phi_+,  \\
\label{eqp}
&&  (\d - H \w )(e^{A -\phi } {\rm Re} \, \Phi_+) =0 \, , \\
\label{eqpb}
&& (\d - H \w )(e^{3 A -\phi } {\rm Im} \, \Phi_+) = -3e^{2A-\phi}{\rm Im} \, (\bar{\mu}\Phi_-) -\frac{1}{8} e^{4 A} \ast \lambda(F) \,, 
\eea
\end{subequations}
 where $\phi$ is the dilaton and $F$  is the sum of the RR field strength on  $M$,  $F = F_1 + F_3 + F_5$ and where $\lambda$ acts on a form as  the transposition of all indices
 \beq
 \label{lambdaop}
 \lambda(\omega_p) = (-)^{\lfloor p/2\rfloor} \omega_p \, .
 \eeq
The ten-dimensional fluxes are defined in terms of $F$ by
\beq
F_{(10)} = {\rm vol}_4 \wedge \lambda (\ast F) + F \, .
\eeq

The complex number $\mu$ determines the size of the AdS$_4$ cosmological constant 
\beq
\Lambda=-3|\mu|^2\,.
\eeq

It is convenient to introduce
 the rescaled forms
 \bea
\hat{\omega} =  e^{i \theta} \omega \, , \\
\hat{z}=\frac{\bar{\mu}}{|\mu|}z \, , 
 \eea
but for simplicity of notation, we  will drop  the $\hat{~}$ symbols in the rest of the paper.

\vspace{0.3cm}

Plugging the explicit form of  \eqref{Phim} and \eqref{Phip},  into the SUSY variations  \eqref{eqm}-\eqref{eqpb}, one can deduce the 
general conditions for AdS$_4$ $\mathcal{N}=1$ SUSY vacua in terms of   the forms  $z$, $\omega$, $j$ and the fluxes. 
As discussed in \cite{Petrini:2013ika},   \eqref{eqm}  implies
\beq
\kpar = 0 \qquad \mbox{or} \qquad \cos \theta  = 0 \, .
\eeq
We will choose the first case namely a strict SU(2) structure. In this case, the equations \eqref{eqm}-\eqref{eqpb} become :
\begin{subequations}
\begin{align}
(d-H\wedge)(e^{3A-\phi}z\wedge e^{-i j})=&2|\mu|e^{2A-\phi}(\omega_I-z_Rz_I\omega_R)\label{eqphi1}\\
(d-H\wedge)(e^{2A-\phi}(\omega_I-z_Rz_I\omega_I))=&0\label{eqphi2}\\
(d-H\wedge)(e^{4A-\phi}(\omega_R+z_Rz_I\omega_I))=&-3|\mu|e^{3A-\phi}\Im(z\wedge e^{ij})+e^{4A}\ast \lambda(F)\label{eqphi3}
\end{align}
\end{subequations}
with $R$ and $I$ denoting the real and imaginary part.

\section{Description of the method}\label{method}

In this section, we present the semi-algorithmical method used to find new sourceless vacua in $\mathcal{N}=1$, AdS$_4$ in type IIB. In fact this method can be extended to all problems where parts of the equations are linear, ie of the type (\ref{lineareq}), and parts of the equations are quadratic/differential, ie of the type (\ref{quadraeq}). 

\subsection{Step 0 : Definitions}

Let $e^i$ be a 6D vielbein on the internal manifold. Define :
\begin{align}
z_1=&e^A(e^1+i e^2) & z_2=&e^A(e^3+i e^4) & z_3=&e^A(e^5+ie^6)
\end{align}
Then
\begin{align}
z=&z_1 \\
 j=&\frac{i}{2}(z_2\wedge \overline{z}_2+z_3\wedge \overline{z}_3)=e^{2A}(e^{34}+e^{56}) &
 \\ \omega=&z_2\wedge z_3=e^{2A}(e^{35}-e^{46}+i(e^{36}+e^{45}))
 \end{align}
define a SU(2) structure on the internal manifold. Moreover define
 \begin{align*}
de^i=&-\frac{1}{2}f^i_{~jk}e^{jk} & dA=&dA_i e^i & d\phi=&d\phi_i e^i & F_1=&F_{1i}e^i\\
F_3=&F_{3i}\omega_3^i & F_5=&F_{5i}\omega_5^i & H=&H_i \omega_3^i
\end{align*}
where $\omega_k^i$ are the canonical real basis of k-forms on a 6-dimensional manifold (for example $\omega_{3i}=\{e^{123}, e^{124},...\}$).

We are looking for a sourceless solution in type IIB with a strict SU(2) structure internal manifold. That is to say that we have to solve for (\ref{eqphi1})-(\ref{eqphi3}) and for the sourceless Bianchi identities $dH=0$ and $d_H F=0$. We will also require that $d(d(e^i))=0$ in order to constrain more the system and be sure to obtain a well-defined manifold at the end of the day.

 We also define the following set of variables:
\begin{align*}
T_i=\left\{f^i_{~jk},dA_i,d\phi_i,F_{1i},F_{3i},F_{5i},H_{3i}\right\}
\end{align*}
to which we will add the parameter $T_0=|\mu|$. It is important to put this parameter with the variables in order to obtain fully linear (\ref{lineareq}) and fully quadratic (\ref{quadraeq}) equations. We can claim we have a solution when we find a set of $T_i$ that solves the aforementioned equations. Indeed, all the equations of motion are solved in this case (see for example \cite{Grana:2006kf} and references therein).

\subsection{Step 1 : Obtaining linear constraints}
The equations (\ref{eqphi1})-(\ref{eqphi3}) are linear in the $T_i$'s and of the form :
\begin{align}
E_{1i}=\left\{\sum_k \alpha^{~k}_i T_k=0\right\}\label{lineareq}
\end{align}
We can easily solve for them and thus eliminate some of the $T_i$'s.
\subsection{Step 2 : Eliminating the derivative in the quadratic equations}
The rest of the equations are quadratic in the $T_i$'s and are of the form :
\begin{align}
E_{2i}=\left\{\sum_{j,k} \alpha_i^{~jk} (dT)_{jk}+\sum_{j,k}\beta_i^{~jk}T_j T_k=0\right\}\label{quadraeq}
\end{align}
where we defined $d(T_i)=(dT)_{ij}e^j$. If these equations are quadratic in the $T_i$'s, they are linear in the $(dT)_{ij}$'s so we can "solve" for them to simplify the system and obtain two sets of equations of the type :
\begin{align}
E_{3i}=&\left\{\sum_{j,k} \alpha_i^{~jk} (dT)_{jk}+\sum_{j,k}\beta_i^{~jk}T_j T_k=0\right\}\label{eqdT}\\
E_{4i}=&\left\{\sum_{kl}\alpha_i^{~kl}T_kT_l=0\right\}\label{eqTT}
\end{align}
Maybe it can be better explained with an example. Assume the system $E_{2i}$ is composed of two equations $(dT)_{12}+T_1T_2=0$ and $(dT)_{12}+(T_2)^2=0$, "solving" for $(dT)_{12}$ means keeping one of the two equations unchanged, and replace $(dT)_{12}$ in the other one to obtain the system $E_{4i}$ : $T_1T_2=(T_2)^2$. In other words, we are splitting $E_{2i}$ in two, one part, $E_{3i}$ with all the $(dT)_{ij}$'s and the other, $E_{4i}$ with only $T_i$'s.

\subsection{Step 3 : Simplifying the leftover quadratic equations}

We can still simplify a bit the system of equations $E_{4i}$ (\ref{eqTT}). Indeed, in general, all the equations are not independent and there exists a simple trick to easily get a minimal system. Simply define $(TT)_{ij}=T_i T_j$, with $i\leq j$. Then the system is linear in these new variables and by solving it, one obtains a minimal system in the $(TT)_{ij}$'s. One just has to go back to the $T_i$'s to have simplified $E_{4i}$. Moreover while solving for the $(TT)_{ij}$'s, we can make it so that $(TT)_{0j}$ appears as much as possible. It will help us get simpler equations for step 4.

\subsection{Step 4 : Adding linear constraints}

The goal of this step is to obtain a linear constraint from the set of quadratic constraints to simplify the original problem. This is inspired by \cite{Solard:2013fva} where some linear conditions were hidden in the quadratic constraints and permitted to fully solve these equations. 

Having simplified the system in steps 2 and 3, some equations may immediately give such a linear constraint:
\begin{itemize}
\item One of the equation can be of the form $\sum_i (\sum_k \alpha_i^k T_k)^2=0$.  Then the linar constraints are $\sum_k \alpha_i^k T_k=0$ for all $i$. 
\item One of the equations can be of the form $T_0 \sum_k \alpha^k T_k=0$. Since $T_0=|\mu|$, which is non-zero since the external manifold is AdS, one can conclude $\sum_k \alpha^k T_k=0$ which is linear.
\end{itemize}
If one is not in one of the case above, one has to make an assumption. The system can often give an hint on what is a sensible assumption or not. 
Indeed, some equations are simpler than other and help make a choice. But how can one find these simpler equations in a system which can be quite complicated? The answer is to look at the eigenvalues of $\alpha^{kl}$ seen as a matrix in \ref{eqTT}. The equations with a small number of non-zero eigenvalues are usually sufficiently simple to make sensible assumptions (see section \ref{LTsol} for an explicit example).

\subsection{Step 5 : Going back to step 1}

We are now going back to step 1 with the additional linear constraints obtained in step 4. We are forced to do all the work again for the following reason. Assume you had for example the equation $(dT)_{11}=(T_2)^2$ in the system $E_{3i}$ (\ref{eqdT}) and that you found $T_1=0$ as a linear constraint in step 4. Then it implies $(dT)_{11}=0$ and so $(T_2)^2=0$ in step 2 which will give the linear constraint $T_2=0$ in step 4. This is the strength of the method : simplify sufficiently the quadratic constraints to spot the linear constraints hidden in them to be able to discover even more linear constraints. 

Thus we are going from step 1 to step 4 to step 1 again until one of the three following things happen :
\begin{itemize}
\item the system has no solution : it means that one of the assumptions made in step 4 is wrong and should be discarded or that there is no solution within the ansatz one was given.
\item $E_{4i}$ (\ref{eqTT}) is empty then one can go to the final step
\item $E_{4i}$ (\ref{eqTT}) is not empty but is sufficiently simple to be able to find a non linear solution of it. Then one can go to the final step.
\end{itemize}

\subsection{Final Step : Solving the last equations}

Ideally at this point both $E_{3i}$ and $E_{4i}$ defined in step 2 are empty but this is often not the case. Nevertheless, they are usually sufficiently simple to be solved by traditional methods. To sum up, the above steps take care of the linear parts of the equations and of some of the quadratic constraints by assuming some linear constraints. What is left are the differential and quadratic parts. An explicit example of this step will be given in section \ref{LTsol}

\section{Examples of new vacua}\label{examples}
In this section we give some examples of solutions found by the above method. One of them is already known as a L\"ust-Tsimpis solution \cite{Lust:2009mb}. The other two, as far as the author knows, are two new vacua in type IIB. 

\subsection{An example of L\"ust-Tsimpis solution}\label{LTsol}

We will give an example of a L\"ust-Tsimpis solution \cite{Lust:2009mb}. In this section we will also give a detailed account of how the method works in this particular case. 

First of all we assume that there is no vector or tensor in the torsion classes as they do in \cite{Lust:2009mb}. These are linear constraints in our variables $T_i$'s so can already be put in step 1. We will also require $de^2=de^3=0$ that is to say, we want $e^2=dx^2$ and $e^3=dx^3$, $x^2$ and $x^3$ being coordinates. This requirement is also linear and can be put in step 1. Finally, we will require that all the variables are functions of only $x^2$ and $x^3$. Part of this requirement is linear (for example, $dA_j=0$ for $j\neq 2,3$). The other part is differential and means that $(dT)_{ij}=0$ for $j\neq 2,3$ and appears in step 2.

We run the algorithm from step 1 to step 3 and take a look at the resulting system $E_{4i}$ (\ref{eqTT}). It contains several simple equations :$|\mu|(f^4_{~15}-\frac{5|\mu|}{2})=|\mu|f^4_{~45}=|\mu| f^4_{~46}=|\mu|(f^5_{~35}-f^4_{~34})=|\mu|f^5_{~45}=0$. We add these linear (since $|\mu|\neq 0$) constraints to step 1.

 Then we rerun the algorithm from step 1. In step 4, we obtain only one equation in $E_{4i}$ namely : $(f^4_{~25})^2+4 (dA_2)^2-\frac{5|\mu|^2}{4}=0$. We are in the case where there is no obvious linear constraint. So we will make a choice : $f^4_{~25}=0$ and $dA_2=\frac{\sqrt{5}|\mu|}{4}$ to solve it.
 
  We rerun the algorithm from step 1 and find that $E_{4i}$ is empty. So we go to the final step and take a look at $E_{3i}$ (\ref{eqdT}). There are 4 equations in it (the projections on $e^2$ and $e^3$ of the two following expressions):
\begin{align}
d(f^{4}_{~34})=&(2 (f^4_{~34})^2-f^4_{~34}f^6_{~36})e^3\\
d(f^{6}_{~36})=&(5|\mu|^2-2 (f^4_{~34})^2+2 f^4_{~34}f^6_{~36}+(f^6_{~36})^2)e^3
\end{align}
There exists a simple solution to this system : $f^4_{~34}=0$ and $f^6_{~36}(x^3)=\sqrt{5}|\mu|\tan(\sqrt{5}|\mu|(x^3-x_0))$ with $x_0$ an
integration constant. With this, $E_{3i}$ and $E_{4i}$ are empty which means we have successfully solved all the relevant equations.

Let's now give explicitly the results. We have :
\begin{align}
de^1=&2|\mu|(e^{36}+e^{45}) & de^2=&d(dx^2)=0 & de^3=&d(dx^3)=0\\\nonumber
de^4=&-\frac{5}{2}|\mu|e^{15}-f^6_{36}e^{56} & de^5=&\frac{5}{2}|\mu|e^{14}+f^6_{36}e^{46} & de^6=&-f^6_{36} e^{36}
\end{align}
The fluxes, dilaton and warp factor being :
\begin{subequations}
\begin{align}
F_1=&0 \\
 F_3=&\frac{|\mu|e^{-2A}}{2}(\sqrt{5}(-e^{135}+e^{146})-e^{234}-e^{256})\\
 F_5=&3|\mu|e^{13456}\\
H=&\frac{|\mu|e^{2A}}{2}(\sqrt{5}(e^{134}+e^{156})+e^{235}-e^{246})\\
\phi=&4A=\sqrt{5}|\mu|x^2 & 
\end{align}
\end{subequations}
with $f^6_{~36}(x^3)=\sqrt{5}|\mu|\tan(\sqrt{5}|\mu|(x^3-x_0))$ with $x_0$ an
integration constant. This is a solution of the $\mathcal{N}=1$ SUSY equations, the sourceless Bianchi identities and $d(d(e^i))=0$ and so of all the equations of motion. Moreover, there are no vectors and no tensors in the torsion classes.

In order to understand more this solution, it is useful to give a coordinate expression of the metric or at least identify each part of the space. In that regard, one can take the following change of variables :
\begin{align}
e^1=&\frac{2\tilde{e}^1}{5|\mu|}+\frac{2f(x^3)dx^6}{5|\mu|\sqrt{5|\mu|^2+f(x^3)^2}} & e^2=&dx^2 & e^3=&dx^3\\
e^4=&\frac{\tilde{e}^4}{\sqrt{5}|\mu|} & e^5=&\frac{\tilde{e}^5}{\sqrt{5}|\mu|} & e^6=&\frac{dx^6}{\sqrt{5|\mu|^2+f(x^3)^2}}\nonumber
\end{align}
with $f(x^3)=\sqrt{5}|\mu|\tan(\sqrt{5}|\mu|(x^3-x_0))$ with $x_0$ an integration constant and  the triplet $\{\tilde{e}^1,\tilde{e}^4,\tilde{e}^5\}$ parametrizing a SU(2) $d\tilde{e}^1=\tilde{e}^{45},~d\tilde{e}^4=-\tilde{e}^{15},~d\tilde{e}^5=\tilde{e}^{14}$. If one wants to see explicitly the squashed Sasaki-Einstein of  L\"ust-Tsimpis \cite{Lust:2009mb}, we now give the correspondence with their objects (note that for us $W_{LT}=|\mu|$ and $c_{LT}=0$) :
\begin{align*}
u_{LT}=&\frac{5|\mu|}{6}e^1\\
dt_{LT}=&e^2\\
\gamma_{LT}=&-\frac{5|\mu|^2}{6}(e^{36}+e^{45})\\
\alpha_{LT}=&\frac{5}{6}|\mu|^2(\sin(\theta_{LT})(e^{34}+e^{56})+\cos(\theta_{LT})(e^{35}-e^{46}))\\
\beta_{LT}=&\frac{5}{6}|\mu|^2(\cos(\theta_{LT})(e^{34}+e^{56})-\sin(\theta_{LT})(e^{35}-e^{46}))
\end{align*}
with $\theta_{LT}$ a constant.
\subsection{A new solution with constant dilaton}

Applying the method to more complex cases, we were able to identify two new vacua. Here we present the first one which has the particularity to have a constant dilaton. We will make an ansatz on the solution to make the method converge more rapidly (this ansatz has been found by trial and error from the general case). We will assume that $de^3=dx_3$ and that all the variables depend on $x_3$ only. We will also assume that $de^2=-f^2_{23}e^{23}$, $de^4=-f^4_{34}e^{34}$ and $de^5=-f^5_{35}e^{35}$. Then some iterations of the algorithm give the following algebra :
\begin{subequations}
\begin{align}
de^1=&(4dA_3(x_3)-d\phi_3(x_3))e^{13}+2|\mu|(e^{36}+e^{45})\\
de^2=&(4dA_3(x_3)-d\phi_3(x_3))e^{23}\\
de^3=&d(dx_3)=0\\
de^4=&-f^4_{34}(x_3)e^{34}\\
de^5=&-(d\phi_3(x_3)+f^4_{34}(x_3))e^{35}\\
de^6=&5|\mu|e^{13}-f^6_{23}(x_3)e^{23}-(2dA_3(x_3)-d\phi_3(x_3)-f^4_{34}(x_3))e^{36}\nonumber\\
&+(4dA_3(x_3)-2d\phi_3(x_3)-2f^4_{34}(x_3))e^{45}
\end{align}
\end{subequations}
with $dA=dA_3(x_3)e^3$ and $d\phi=d\phi_3(x_3) e^3$. Moreover, in order to verify the Bianchi identities and $d(d(e^i))=0$, the four functions verify the following equations (which are the system $E_{3i}$ (\ref{eqdT}) in this case) :
\begin{subequations}
\begin{align}
(dA_3)'=&\frac{5|\mu|^2}{2}+6(dA_3)^2+dA_3 f^4_{34}+\frac{1}{2}(f^6_{23})^2\\
(d\phi_3)'=&10 dA_3 d\phi_3-2(d\phi_3)^2+d\phi_3 f^4_{34}+(f^6_{23})^2\\
(f^4_{34})'=&10|\mu|^2+16(dA_3)^2-16dA_3d\phi_3+4(d\phi_3)^2-6dA_3f^4_{34}+4d\phi_3f^4_{34}+3(f^4_{34})^2\\
(f^6_{23})'=&4dA_3 f^6_{23}
\end{align}
\end{subequations}
Unfortunately, the author hasn't been able to solve these equations in all generality. But there exists the following more simple solution (which is the above one with $d\phi=0$, $f^4_{34}=4dA_3$ and $f^6_{23}=0$):
\begin{subequations}
\begin{align}
de^1=&4dA_3(x_3)e^{13}+2|\mu|(e^{36}+e^{45})\\
de^2=&4dA_3(x_3)e^{23}\\
de^3=&d(dx_3)=0\\
de^4=&-4dA_3(x_3)e^{34}\\
de^5=&-4dA_3(x_3)e^{35}\\
de^6=&5|\mu|e^{13}+2dA_3e^{36}-4dA_3e^{45}
\end{align}
\end{subequations}
The fluxes, dilaton and warp factor being :
\begin{subequations}
\begin{align}
F_1=&0\\
F_3=&e^{2A}(4dA_3(x_3) e^{125}+|\mu|(-3e^{234}+2e^{256}))\\
F_5=&3e^{4A}|\mu| e^{13456}\\
H=&e^{2A}(4dA_3(x_3)e^{124}+|\mu|(3e^{235}+2e^{246}))\\
\phi=&0\\
dA_3(x_3)=&\frac{|\mu|}{2}\tan(5|\mu|(x_3-x_0))\\
A(x_3)=&-\frac{1}{10}\log(\cos(5|\mu|(x_3-x_0)))
\end{align}
\end{subequations}
 with $x_0$ an integration constant.  This is a solution of the $\mathcal{N}=1$ SUSY equations, the sourceless Bianchi identities and $d(d(e^i))=0$ and so of all the equations of motion.
 
 We then put its expression in coordinates by the following change of variables :
 \begin{subequations}
 \begin{align}
 e^1=&2|\mu|f(x_3)^{\frac{2}{5}}\tilde{e}^1+\frac{2|\mu|}{dA_3(x_3)f(x_3)^{\frac{3}{10}}}dx^6\\
 e^2=&f(x_3)^{\frac{1}{5}}dx^2\\
 e^3=&dx_3=dA_3(x_3)f(x_3)^{\frac{17}{10}}dx'_3\\
 e^4=&f(x_3)^{\frac{1}{5}}dx^4\\
 e^5=&f(x_3)^{\frac{1}{5}}dx^5\\
 e^6=&4dA_3(x_3)f(x_3)^{\frac{2}{5}}\tilde{e}^1+\frac{10}{f(x_3)^{\frac{3}{10}}}dx^6
 \end{align}
 \end{subequations}
 with $d\tilde{e}^1=dx'_3\wedge dx^6+dx^4\wedge dx^5$ and $f(x_3)=\frac{1}{\frac{5|\mu|^2}{2}+10 dA_3(x_3)^2}$. Unfortunately, the author has not been able to obtain an explicit change of variables to go from $x_3$ to $x'_3$.

\subsection{A new solution with non constant dilaton}
\subsubsection{The solution}
Another solution arose from the method described, one with non constant dilaton. Once again to make the method converge more rapidly one takes an ansatz (this ansatz has been found by trial and error from the general case).  We will assume that $de^3=dx_3$ and that all the variables depend on $x_3$ only. We will also assume that  $de^4=-f^4_{34}e^{34}$ and $H=0$. After some iterations of the algorithm, one obtains :
\begin{subequations}
\begin{align}
de^1=&dA_3(x_3)e^{13}+2|\mu|(e^{36}+e^{45})\label{algsol31}\\
de^2=&dA_3(x_3) e^{23}\label{algsol32}\\
de^3=&d(dx_3)=0\label{algsol33}\\
de^4=&\frac{|\mu|^2}{dA_3(x_3)}e^{34}\label{algsol34}\\
de^5=&2|\mu|e^{14}-dA_3(x_3)e^{35}-\frac{|\mu|^2}{dA_3(x_3)}e^{46}\label{algsol35}\\
de^6=&2|\mu|e^{13}-dA_3(x_3)e^{36}+\frac{|\mu|^2}{dA_3(x_3)}e^{45}\label{algsol36}
\end{align}
\end{subequations}
The fluxes, dilaton and warp factor being :
\begin{subequations}
\begin{align}
F_1=&e^{-\phi}(|\mu| e^1-2dA_3(x_3)e^6)\\
F_3=&e^{2A-\phi}(dA_3(x_3)e^{125}-3|\mu| e^{234}+|\mu| e^{256})\\
F_5=&3|\mu| e^{13456}\\
H=&0\\
dA=&dA_3(x_3)e^3=\frac{|\mu|}{2}\tan\left(2|\mu|(x_3-x_0)\right)e^3\\
A=&-\frac{1}{4}\log\left(\cos\left(2|\mu|(x_3-x_0)\right)\right)
\\
d\phi=&d\phi_3 e^3=3dA_3(x_3) e^3\\
\phi=&3A+cst
\end{align}
\end{subequations}
with $x_0$ an integration constant. This is a solution of the $\mathcal{N}=1$ SUSY equations, the sourceless Bianchi identities and $d(d(e^i))=0$ and so of all the equations of motion.

Once again a coordinate expression is useful. Do the following change of variables :
\begin{subequations}
\begin{align}
e^1=&\cos(2|\mu|x_4)\cos(X)\sin(X)^{\frac{1}{4}}dx_1+\sin(2|\mu|x_4)\cos(X)\sin(X)^{\frac{1}{4}}dx_5+\sin(X)^{\frac{5}{4}}dx_6\\
e^2=&\sin(X)^{\frac{1}{4}}dx_2\\
e^3=&\frac{dX}{2|\mu|}\\
e^4=&\cos(X)dx_4\\
e^5=&-\sin(2|\mu|x_4)\sin(X)^{\frac{1}{4}}dx_1+\cos(2|\mu|x_4)\sin(X)^{\frac{1}{4}}dx_5\\
e^6=&-\cos(2|\mu|x_4)\sin(X)^{\frac{5}{4}}dx_1-\sin(2|\mu|x_4\sin(X)^{\frac{5}{4}}dx_5+\cos(X)\sin(X)^{\frac{1}{4}}dx_6
\end{align}
\end{subequations}
with  $X=(2|\mu|(x_3-x_0))+\frac{\pi}{2}$. Note that $e^A=\sin(X)^{-\frac{1}{4}}$. We give the expression of the metric in the $(x_1,x_2,X,x_4,x_5,x_6)$ system of coordinates :
\begin{align}
 g_{ij}=\left(
 \begin{array}{cccccc}
1 &0 & 0 &0 & 0& 0\\
  0& 1 & 0 &0 & 0& 0\\
   0& 0 & \frac{1}{4|\mu|^2\sqrt{\sin(X)}}&0 & 0& 0\\
    0& 0 & 0 &\frac{\cos(X)^2}{\sqrt{\sin(X)}}& 0& 0\\
     0& 0 & 0 &0 & 1& 0\\
      0& 0 & 0 &0 & 0& 1
 \end{array} \right)\label{met3}
 \end{align}
One can also calculate the Ricci scalar : $R=|\mu|^2\frac{1-3\cos(2X)}{(\sin(X))^{\frac{3}{2}}}$ which goes to infinity when $X$ goes to 0. This shows that, a priori, this space is singular. Around 0, this metric doesn't have the form of the D-brane metric so one has to better understand this singularity. In order to do that, let's look at the ten dimensional metric around $X=0$ at first order:
\begin{align}
ds^2=\frac{1}{\sqrt{X}}ds^2_{(4)}+(dx_1^2+dx_2^2+dx_5^2+dx_6^2)+\frac{dX^2}{4|\mu|^2\sqrt{X}}+\frac{dx_4^2}{\sqrt{X}}
\end{align}
Then define $\tilde{x}=\frac{\sqrt{X}}{|\mu|}$, the metric around 0 becomes :
\begin{align}
ds^2=\frac{1}{|\mu|\tilde{x}}(ds^2_{(4)}+dx_4^2)+(dx_1^2+dx_2^2+dx_5^2+dx_6^2)+|\mu|\tilde{x}d\tilde{x}^2
\end{align}
This shows that this system can be mapped to a D5-D7 intersecting system which are delocalized in the \{1,2,5,6\} directions. For example D5 along $x^1, x^4$ and D7 along $x^2, x^4, x^5, x^6$. Indeed, we are in the case of a system similar to (10) of \cite{Gauntlett:1996pb} with only one transverse direction for both branes (the $\tilde{x}$ direction), and $H_5=H_7=|\mu|\tilde{x}$ being the associated harmonic function. Similarly according to equation (478) of \cite{Youm:1997hw}, one has $e^{-2\phi}=H_5H_7^{2}=|\mu|^3\tilde{x}^3$ which corresponds to the dilaton value on the solution around X equal zero : $e^{-2\phi}=e^{-6A}=\sin(X)^{\frac{3}{2}}=_{X\rightarrow 0}X^{\frac{3}{2}}=|\mu|^3\tilde{x}^3$.
\subsubsection{T-dual solution}

One can see that there exists several isometric directions for this solution (at first sight $dx_1$, $dx_2$, $dx_5$, $dx_6$). To illustrate this, we will explicitly give the T-dual along the $dx_2=e^A e^2$ direction. The resulting solution in IIA is :
\begin{subequations}
\begin{align}
de^1=&dA_3(x_3)e^{13}+2|\mu|(e^{36}+e^{45})\\
de^2=&dA_3(x_3) e^{23}\\
de^3=&d(dx_3)=0\\
de^4=&\frac{|\mu|^2}{dA_3(x_3)}e^{34}\\
de^5=&2|\mu|e^{14}-dA_3(x_3)e^{35}-\frac{|\mu|^2}{dA_3(x_3)}e^{46}\\
de^6=&2|\mu|e^{13}-dA_3(x_3)e^{36}+\frac{|\mu|^2}{dA_3(x_3)}e^{45}
\end{align}
\end{subequations}
The fluxes, dilaton and warp factor being :
\begin{subequations}
\begin{align}
F_0=&F_4=H=0\\
F_2=&e^{A-\phi}(|\mu|e^{12}-2dA_3e^{15}+2dA_3e^{26}-3|\mu|e^{34}+|\mu|e^{56})\\
F_6=&3e^{5A-\phi}|\mu|e^{123456}\\
dA=&dA_3(x_3)e^3=\frac{|\mu|}{2}\tan\left(2|\mu|(x_3-x_0)\right)e^3\\
A=&-\frac{1}{4}\log\left(\cos\left(2|\mu|(x_3-x_0)\right)\right)
\\
d\phi=&d\phi_3 e^3=3dA_3(x_3) e^3\\
\phi=&3A+cst
\end{align}
\end{subequations}
Note that the space the solution lives on is the same in both IIA and IIB. But in IIA, contrary to IIB, we have, the following SU(3) structure :
\begin{align}
z_1=&e^A(i e^1- e^2) & z_2=&e^A(e^3+i e^4) & z_3=&e^A(i e^5-e^6)
\end{align}
\begin{subequations}
\begin{align}
z=&z_1 \\
 j=&\frac{i}{2}(z_2\wedge \overline{z}_2+z_3\wedge \overline{z}_3)=e^{2A}(e^{34}+e^{56}) &
 \\ \omega=&z_2\wedge z_3=e^{2A}(-e^{35}+e^{46}+i(-e^{36}-e^{45}))\\
 \Omega=&z\wedge \omega\\
 J=&\frac{i}{2}z\wedge\bar{z}+j\\
 \Phi_+=&-\frac{ie^A}{8}e^{-iJ}\\
 \Phi_-=&-\frac{i e^A}{8}\Omega
 \end{align}
 \end{subequations}

\section*{Conclusion and outlooks}

In this paper, we managed to identify two new  vacua in type IIB which are explicit. It is a step forward in identifying the web of vacua in type II. We also have been able to discover a new IIA solution by applying T-duality along an isometric direction on one of the solutions. One caveat should be pointed out : these solutions are indeed sourceless if the space is smooth which is not guaranteed by the analysis. Indeed, one could find localized sources (or partially localized sources as we did for the second example) but it is not in the scope of this paper.

 To obtain these new vacua, we devised a semi-algorithmical method which can be applied to lots of other similar situations. Indeed, one can apply it to type IIA to discover new vacua (and we should be able to easily recover the one we found here), or to type IIB with dynamic SU(2) structure instead of the strict SU(2) structure we restricted to in this paper.  More generaly, one can apply it to all problems with a linear part and a quadratic/differential part of the type (\ref{lineareq},\ref{quadraeq}). In that respect, one can see this paper as a proof of concept for the method.
 
 There is also lots of room for improvement for the method depending on which problems one applies it to. Indeed, in this paper we restricted to having only one parameter which had to be non zero $|\mu|$. In fact it is quite common to find other linear combinations of variables to be non zero. Then one can modify step 3 and step 4 to take that into account and be provided with even more linear constraints. Another improvement concerns the automatization. In step 4, it is quite common to have constraints of the type  $(\sum_i\alpha^i T_i)(\sum\beta^i T_i)=0$. One can incorporate this case in the algorithm to build a tree of assumptions (here one branch is given by $(\sum_i\alpha^i T_i)=0$ and the other by $(\sum\beta^i T_i)=0$) instead of just choosing one.

\section*{Acknowledgement}
The author would like to thank Joohno Kim, Dario Rosa and Alessandro Tomasiello for useful discussions . The author is supported in part by INFN and by the European Research Council under the European Union's Seventh Framework Program (FP/2007-2013) ERC Grant Agreement n.307286 (XD-STRING).

\bibliographystyle{utphysmodb}
\bibliography{SUSYAdS}

\providecommand{\href}[2]{#2}\begingroup\raggedright\begin{thebibliography}{10}

\bibitem{Tomasiello:2007eq}
A.~Tomasiello,  {\em {New string vacua from twistor spaces}}, Phys.Rev. {\bf
  D78} (2008) 046007
[\href{http://www.arXiv.org/abs/0712.1396}{{\tt 0712.1396}}].
%%CITATION = ARXIV:0712.1396;%%.

\bibitem{Petrini:2009ur}
M.~Petrini and A.~Zaffaroni,  {\em {N=2 solutions of massive type IIA and their
  Chern-Simons duals}}, JHEP {\bf 09} (2009) 107
[\href{http://www.arXiv.org/abs/0904.4915}{{\tt 0904.4915}}].
%%CITATION = ARXIV:0904.4915;%%.

\bibitem{Lust:2009mb}
D.~Lust and D.~Tsimpis,  {\em {New supersymmetric AdS(4) type II vacua}}, JHEP
  {\bf 0909} (2009) 098
[\href{http://www.arXiv.org/abs/0906.2561}{{\tt 0906.2561}}].
%%CITATION = ARXIV:0906.2561;%%.

\bibitem{Koerber:2010rn}
P.~Koerber and S.~Kors,  {\em {A landscape of non-supersymmetric AdS vacua on
  coset manifolds}}, Phys.Rev. {\bf D81} (2010) 105006
[\href{http://www.arXiv.org/abs/1001.0003}{{\tt 1001.0003}}].
%%CITATION = ARXIV:1001.0003;%%.

\bibitem{Guarino:2015jca}
A.~Guarino, D.~L. Jafferis and O.~Varela,  {\em {String Theory Origin of Dyonic
  N=8 Supergravity and Its Chern-Simons Duals}}, Phys. Rev. Lett. {\bf 115}
  (2015), no.~9, 091601
[\href{http://www.arXiv.org/abs/1504.08009}{{\tt 1504.08009}}].
%%CITATION = ARXIV:1504.08009;%%.

\bibitem{Dasgupta:1999ss}
K.~Dasgupta, G.~Rajesh and S.~Sethi,  {\em {M theory, orientifolds and G -
  flux}}, JHEP {\bf 9908} (1999) 023
[\href{http://www.arXiv.org/abs/hep-th/9908088}{{\tt hep-th/9908088}}].
%%CITATION = HEP-TH/9908088;%%.

\bibitem{Giddings:2001yu}
S.~B. Giddings, S.~Kachru and J.~Polchinski,  {\em {Hierarchies from fluxes in
  string compactifications}}, Phys.Rev. {\bf D66} (2002) 106006
[\href{http://www.arXiv.org/abs/hep-th/0105097}{{\tt hep-th/0105097}}].
%%CITATION = HEP-TH/0105097;%%.

\bibitem{Behrndt:2004km}
K.~Behrndt and M.~Cvetic,  {\em {General N = 1 supersymmetric flux vacua of
  (massive) type IIA string theory}}, Phys.Rev.Lett. {\bf 95} (2005) 021601
[\href{http://www.arXiv.org/abs/hep-th/0403049}{{\tt hep-th/0403049}}].
%%CITATION = HEP-TH/0403049;%%.

\bibitem{Derendinger:2004jn}
J.-P. Derendinger, C.~Kounnas, P.~M. Petropoulos and F.~Zwirner,  {\em
  {Superpotentials in IIA compactifications with general fluxes}}, Nucl.Phys.
  {\bf B715} (2005) 211--233
[\href{http://www.arXiv.org/abs/hep-th/0411276}{{\tt hep-th/0411276}}].
%%CITATION = HEP-TH/0411276;%%.

\bibitem{Lust:2004ig}
D.~Lust and D.~Tsimpis,  {\em {Supersymmetric AdS(4) compactifications of IIA
  supergravity}}, JHEP {\bf 0502} (2005) 027
[\href{http://www.arXiv.org/abs/hep-th/0412250}{{\tt hep-th/0412250}}].
%%CITATION = HEP-TH/0412250;%%.

\bibitem{DeWolfe:2005uu}
O.~DeWolfe, A.~Giryavets, S.~Kachru and W.~Taylor,  {\em {Type IIA moduli
  stabilization}}, JHEP {\bf 0507} (2005) 066
[\href{http://www.arXiv.org/abs/hep-th/0505160}{{\tt hep-th/0505160}}].
%%CITATION = HEP-TH/0505160;%%.

\bibitem{Koerber:2008rx}
P.~Koerber, D.~Lust and D.~Tsimpis,  {\em {Type IIA AdS(4) compactifications on
  cosets, interpolations and domain walls}}, JHEP {\bf 0807} (2008) 017
[\href{http://www.arXiv.org/abs/0804.0614}{{\tt 0804.0614}}].
%%CITATION = ARXIV:0804.0614;%%.

\bibitem{Caviezel:2008ik}
C.~Caviezel, P.~Koerber, S.~Kors, D.~Lust, D.~Tsimpis {\em et al.},  {\em {The
  Effective theory of type IIA AdS(4) compactifications on nilmanifolds and
  cosets}}, Class.Quant.Grav. {\bf 26} (2009) 025014
[\href{http://www.arXiv.org/abs/0806.3458}{{\tt 0806.3458}}].
%%CITATION = ARXIV:0806.3458;%%.

\bibitem{Varela:2015uca}
O.~Varela,  {\em {AdS$_{4}$ solutions of massive IIA from dyonic ISO(7)
  supergravity}}, JHEP {\bf 03} (2016) 071
[\href{http://www.arXiv.org/abs/1509.07117}{{\tt 1509.07117}}].
%%CITATION = ARXIV:1509.07117;%%.

\bibitem{Lust:2009zb}
D.~Lust and D.~Tsimpis,  {\em {Classes of AdS(4) type IIA/IIB compactifications
  with SU(3) x SU(3) structure}}, JHEP {\bf 0904} (2009) 111
[\href{http://www.arXiv.org/abs/0901.4474}{{\tt 0901.4474}}].
%%CITATION = ARXIV:0901.4474;%%.

\bibitem{Petrini:2013ika}
M.~Petrini, G.~Solard and T.~Van~Riet,  {\em AdS vacua with scale separation
  from IIB supergravity}, JHEP {\bf 2013} (2013), no.~11, 1--41
[\href{http://www.arXiv.org/abs/1308.1265}{{\tt 1308.1265}}].
%%CITATION = ARXIV:1308.1265;%%.

\bibitem{D'Hoker:2007xy}
E.~D'Hoker, J.~Estes and M.~Gutperle,  {\em {Exact half-BPS Type IIB interface
  solutions. I. Local solution and supersymmetric Janus}}, JHEP {\bf 06} (2007)
  021
[\href{http://www.arXiv.org/abs/0705.0022}{{\tt 0705.0022}}].
%%CITATION = ARXIV:0705.0022;%%.

\bibitem{Assel:2011xz}
B.~Assel, C.~Bachas, J.~Estes and J.~Gomis,  {\em {Holographic Duals of D=3 N=4
  Superconformal Field Theories}}, JHEP {\bf 08} (2011) 087
[\href{http://www.arXiv.org/abs/1106.4253}{{\tt 1106.4253}}].
%%CITATION = ARXIV:1106.4253;%%.

\bibitem{Grana:2006kf}
M.~Grana, R.~Minasian, M.~Petrini and A.~Tomasiello,  {\em {A Scan for new N=1
  vacua on twisted tori}}, JHEP {\bf 0705} (2007) 031
[\href{http://www.arXiv.org/abs/hep-th/0609124}{{\tt hep-th/0609124}}].
%%CITATION = HEP-TH/0609124;%%.

\bibitem{Grana:2004bg}
M.~Grana, R.~Minasian, M.~Petrini and A.~Tomasiello,  {\em {Supersymmetric
  backgrounds from generalized Calabi-Yau manifolds}}, JHEP {\bf 0408} (2004)
  046
[\href{http://www.arXiv.org/abs/hep-th/0406137}{{\tt hep-th/0406137}}].
%%CITATION = HEP-TH/0406137;%%.

\bibitem{Grana:2005sn}
M.~Grana, R.~Minasian, M.~Petrini and A.~Tomasiello,  {\em {Generalized
  structures of N=1 vacua}}, JHEP {\bf 0511} (2005) 020
[\href{http://www.arXiv.org/abs/hep-th/0505212}{{\tt hep-th/0505212}}].
%%CITATION = HEP-TH/0505212;%%.

\bibitem{Gauntlett:2005ww}
J.~P. Gauntlett, D.~Martelli, J.~Sparks and D.~Waldram,  {\em {Supersymmetric
  AdS(5) solutions of type IIB supergravity}}, Class. Quant. Grav. {\bf 23}
  (2006) 4693--4718
[\href{http://www.arXiv.org/abs/hep-th/0510125}{{\tt hep-th/0510125}}].
%%CITATION = HEP-TH/0510125;%%.

\bibitem{Koerber:2007hd}
P.~Koerber and D.~Tsimpis,  {\em {Supersymmetric sources, integrability and
  generalized-structure compactifications}}, JHEP {\bf 0708} (2007) 082
[\href{http://www.arXiv.org/abs/0706.1244}{{\tt 0706.1244}}].
%%CITATION = ARXIV:0706.1244;%%.

\bibitem{Solard:2013fva}
G.~Solard,  {\em {N=1 SUSY $AdS_4$ vacua in IIB SUGRA on group manifolds}},
  JHEP {\bf 02} (2014) 017
[\href{http://www.arXiv.org/abs/1310.4836}{{\tt 1310.4836}}].
%%CITATION = ARXIV:1310.4836;%%.

\bibitem{Behrndt:2005bv}
K.~Behrndt, M.~Cvetic and P.~Gao,  {\em {General type IIB fluxes with SU(3)
  structures}}, Nucl.Phys. {\bf B721} (2005) 287--308
[\href{http://www.arXiv.org/abs/hep-th/0502154}{{\tt hep-th/0502154}}].
%%CITATION = HEP-TH/0502154;%%.

\bibitem{Hitchin:2004ut}
N.~Hitchin,  {\em {Generalized Calabi-Yau manifolds}}, Quart.J.Math.Oxford Ser.
  {\bf 54} (2003) 281--308
[\href{http://www.arXiv.org/abs/math/0209099}{{\tt math/0209099}}].
%%CITATION = MATH/0209099;%%.

\bibitem{Gualtieri:2003dx}
M.~Gualtieri,  {\em {Generalized complex geometry}},
\href{http://www.arXiv.org/abs/math/0401221}{{\tt math/0401221}}.
%%CITATION = MATH/0401221;%%.

\bibitem{Andriot:2008va}
D.~Andriot,  {\em {New supersymmetric flux vacua with intermediate SU(2)
  structure}}, JHEP {\bf 0808} (2008) 096
[\href{http://www.arXiv.org/abs/0804.1769}{{\tt 0804.1769}}].
%%CITATION = ARXIV:0804.1769;%%.

\bibitem{Gauntlett:1996pb}
J.~P. Gauntlett, D.~A. Kastor and J.~H. Traschen,  {\em {Overlapping branes in
  M theory}}, Nucl. Phys. {\bf B478} (1996) 544--560
[\href{http://www.arXiv.org/abs/hep-th/9604179}{{\tt hep-th/9604179}}].
%%CITATION = HEP-TH/9604179;%%.

\bibitem{Youm:1997hw}
D.~Youm,  {\em {Black holes and solitons in string theory}}, Phys. Rept. {\bf
  316} (1999) 1--232
[\href{http://www.arXiv.org/abs/hep-th/9710046}{{\tt hep-th/9710046}}].
%%CITATION = HEP-TH/9710046;%%.

\end{thebibliography}\endgroup
\end{document}